\documentclass[twocolumn]{aastex}
\usepackage{amsmath,esint}
\usepackage{url,aasmacros}
\usepackage{natbib}
\usepackage{soul, CJK}

\usepackage{bm}
\usepackage{xspace}
\usepackage{color}
\usepackage{enumitem}

\newcommand{\degsq}{deg$^2$\ }
\usepackage[utf8]{inputenc}

\shorttitle{DECam Follow-up of S190510g}
\shortauthors{Andreoni \& Goldstein and the GROWTH Collaboration}

\begin{document}

\title{	
GROWTH on S190510g: DECam Observation Planning and Follow-Up of a Distant Binary Neutron Star Merger Candidate
}

\author[0000-0002-8977-1498]{Igor~Andreoni}
\email{andreoni@caltech.edu}
\affil{Division of Physics, Mathematics, and Astronomy, California Institute of Technology, Pasadena, CA 91125, USA}
\author[0000-0003-3461-8661]{Daniel~A.~Goldstein}
\altaffiliation{Hubble Fellow}
\affil{Division of Physics, Mathematics, and Astronomy, California Institute of Technology, Pasadena, CA 91125, USA}
\collaboration{these authors contributed equally to this work}

\author{Shreya Anand}
\affil{Division of Physics, Mathematics, and Astronomy, California Institute of Technology, Pasadena, CA 91125, USA}

\author[0000-0002-8262-2924]{Michael W. Coughlin}
\affil{Division of Physics, Mathematics, and Astronomy, California Institute of Technology, Pasadena, CA 91125, USA}

\author[0000-0001-9898-5597]{Leo P. Singer}
\affiliation{Astrophysics Science Division, NASA Goddard Space Flight Center, MC 661, Greenbelt, MD 20771, USA}
\affiliation{Joint Space-Science Institute, University of Maryland, College Park, MD 20742, USA}

\author[0000-0002-2184-6430]{Tom{\'a}s Ahumada}
\affil{Department of Astronomy, University of Maryland, College Park, MD 20742, USA}

\author[0000-0002-7226-0659]{Michael Medford}
\affiliation{Computational Science Department, Lawrence Berkeley National Laboratory, 1 Cyclotron Road, MS 50B-4206, Berkeley, CA 94720, USA}
\affiliation{Department of Astronomy, University of California, Berkeley, CA 94720-3411, USA}

\author[0000-0002-7252-3877]{Erik C. Kool}
\affiliation{The Oskar Klein Centre \& Department of Astronomy, Stockholm University, AlbaNova, SE-106 91 Stockholm, Sweden}

\author[0000-0003-2601-1472]{Sara Webb}
\affiliation{Centre for Astrophysics and Supercomputing, Swinburne University of Technology, Hawthorn, VIC, 3122, Australia}
\affiliation{ARC Centre of Excellence for Gravitational Wave Discovery (OzGrav), Australia}

\author[0000-0002-8255-5127]{Mattia Bulla}
\affiliation{The Oskar Klein Centre, Department of Physics, Stockholm University, AlbaNova, SE-106 91 Stockholm, Sweden}

\author[0000-0002-7777-216X]{Joshua S. Bloom}
\affiliation{Department of Astronomy, University of California, Berkeley, CA 94720-3411, USA}
\affiliation{Lawrence Berkeley National Laboratory, 1 Cyclotron Road, MS 50B-4206, Berkeley, CA 94720, USA}

\author{Mansi M. Kasliwal}
\affiliation{Division of Physics, Mathematics, and Astronomy, California Institute of Technology, Pasadena, CA 91125, USA}

\author[0000-0002-3389-0586]{Peter E. Nugent}
\affiliation{Computational Science Department, Lawrence Berkeley National Laboratory, 1 Cyclotron Road, MS 50B-4206, Berkeley, CA 94720, USA}
\affiliation{Department of Astronomy, University of California, Berkeley, CA 94720-3411, USA}

\author{Ashot Bagdasaryan}
\affil{Division of Physics, Mathematics, and Astronomy, California Institute of Technology, Pasadena, CA 91125, USA}

\author{Jennifer Barnes}
\altaffiliation{Einstein Fellow}
\affiliation{Columbia Astrophysics Laboratory, Columbia University, New York, NY 10032, USA}

\author[0000-0002-6877-7655]{David O. Cook}
\affiliation{IPAC, California Institute of Technology, 1200 E. California Blvd, Pasadena, CA 91125, USA}

\author[0000-0001-5703-2108]{Jeff Cooke}
\affiliation{Australian Research Council Centre of Excellence for Gravitational Wave Discovery (OzGrav), Hawthorn, VIC, 3122, Australia}
\affiliation{Centre for Astrophysics and Supercomputing, Swinburne University of Technology, Hawthorn, VIC, 3122, Australia}

\author[0000-0001-5060-8733]{Dmitry~A. Duev}
\affil{Division of Physics, Mathematics, and Astronomy, California Institute of Technology, Pasadena, CA 91125, USA}

\author{U.~Christoffer Fremling}
\affiliation{Division of Physics, Mathematics, and Astronomy, California Institute of Technology, Pasadena, CA 91125, USA}

\author[0000-0002-1955-2230]{Pradip Gatkine}
\affil{Department of Astronomy, University of Maryland, College Park, MD 20742, USA}

\author[0000-0001-8205-2506]{V. Zach Golkhou}
\altaffiliation{Moore-Sloan, WRF, and DIRAC Fellow}
\affiliation{DIRAC Institute, Department of Astronomy, University of Washington, 3910 15th Avenue NE, Seattle, WA 98195, USA}
\affiliation{The eScience Institute, University of Washington, Seattle, WA 98195, USA}

\author[0000-0002-5105-344X]{Albert K. H. Kong}
\affiliation{Institute of Astronomy, National Tsing Hua University, Hsinchu 30013, Taiwan}

\author[0000-0003-2242-0244]{Ashish Mahabal}
\affiliation{Division of Physics, Mathematics, and Astronomy, California Institute of Technology, Pasadena, CA 91125, USA}
\affiliation{Center for Data Driven Discovery, California Institute of Technology, Pasadena, CA 91125, USA}

\author{Jorge Mart\'inez-Palomera}
\affiliation{Department of Astronomy, University of California, Berkeley, CA 94720-3411, USA}

\author{Duo Tao}
\affiliation{Division of Physics, Mathematics, and Astronomy, California Institute of Technology, Pasadena, CA 91125, USA}

\author[0000-0002-9870-5695]{Keming Zhang \begin{CJK*}{UTF8}{gkai}(张可名)\end{CJK*}}
\affiliation{Department of Astronomy, University of California, Berkeley, CA 94720-3411, USA}

\begin{abstract}
The first two months of the third Advanced LIGO and Virgo observing run (2019 April--May)  showed that distant gravitational wave (GW) events can now be readily detected. Three candidate mergers containing neutron stars (NS) were reported in a span of 15 days, all likely located more than 100\,Mpc away.
However, distant events such as the three new NS mergers are likely to be coarsely localized, which highlights the importance of facilities and scheduling systems that enable deep observations over hundreds to thousands of square degrees to detect the electromagnetic counterparts.
On 2019-05-10 02:59:39.292 UT the GW candidate S190510g was discovered and initially classified as a BNS merger with 98\% probability. The GW event was localized within an area of 3462\,deg$^2$, later refined to 1166\,deg$^2$ (90\%) at a distance of $227 \pm 92$\,Mpc.  We triggered Target of Opportunity observations with the Dark Energy Camera (DECam), a wide-field optical imager mounted at the prime focus of the 4m Blanco Telescope at CTIO in Chile. This Letter describes our DECam observations and our real-time analysis results, focusing in particular on the design and implementation of the observing strategy.  Within 24 hours of the merger time, we observed 65\% of the total enclosed probability of the final skymap with an observing efficiency of 94\%. We identified and publicly announced 13 candidate counterparts. S190510g was re-classified 1.7 days after the merger, after our observations were completed, with a ``binary neutron star merger” probability reduced from 98\% to 42\% in favor of a ``terrestrial” classification.
\end{abstract}

\section{Introduction}
\label{sec: intro}

The joint detection of electromagnetic (EM) and gravitational wave (GW) signals from the binary neutron star (BNS) merger GW170817 \citep{Abbott2017GW170817discovery} was a watershed moment for astronomy. 
The  discovery of the GW event triggered an extensive EM follow-up campaign, and the resulting panchromatic dataset exacted stringent constraints on fundamental physics \citep{Abbott2017GW170817discovery}, gave new insight into the origin of the heavy elements \citep[e.g.,][]{Arcavi2017GW,Coulter2017,Chornock2017,Cowperthwaite2017,Kasliwal2017,Pian2017Nat, Smartt2017, Kasen2017Nat},  demonstrated a novel technique for measuring cosmological parameters \citep{Abbott2017cosmology}, and marked the beginning of the ``GW multi-messenger era'' \citep{Abbott2017MMA}.

The Swope Supernova Survey first reported the optical counterpart to GW170817 in NGC\,4993 from a galaxy-targeted search \citep{Coulter2017} with independent confirmation by several other teams shortly after \citep{Valenti2017,Arcavi2017GW,Tanvir2017,Lipunov2017,Soares-Santos2017}.
Galaxy-targeted searches pre-select galaxies that could harbor counterparts based on criteria such as sky location, distance, star formation rate, or stellar mass, then search those galaxies for transients.
As galaxy-targeted searches do not require observations of large swaths of sky, they can be carried out on telescopes with small fields-of-view \citep[see for example][]{2018ApJ...857...81G}. 
The galaxy-targeted approach worked particularly well in case of GW170817 because NGC 4993 is relatively nearby ($D = 41.0 \pm 3.1$\,Mpc; \citealt{Hjorth2017ngc4993}, but see also \citealt{Im2017ApJ,Pan2017,Levan2017,Cantiello2018ngc4993}) and  the optical counterpart to GW170817 was bright enough $(M_V \sim -16)$ to be detected with 1m-class telescopes.

\begin{figure*}[htbp!]
    \centering
        \includegraphics[width=0.82\textwidth]{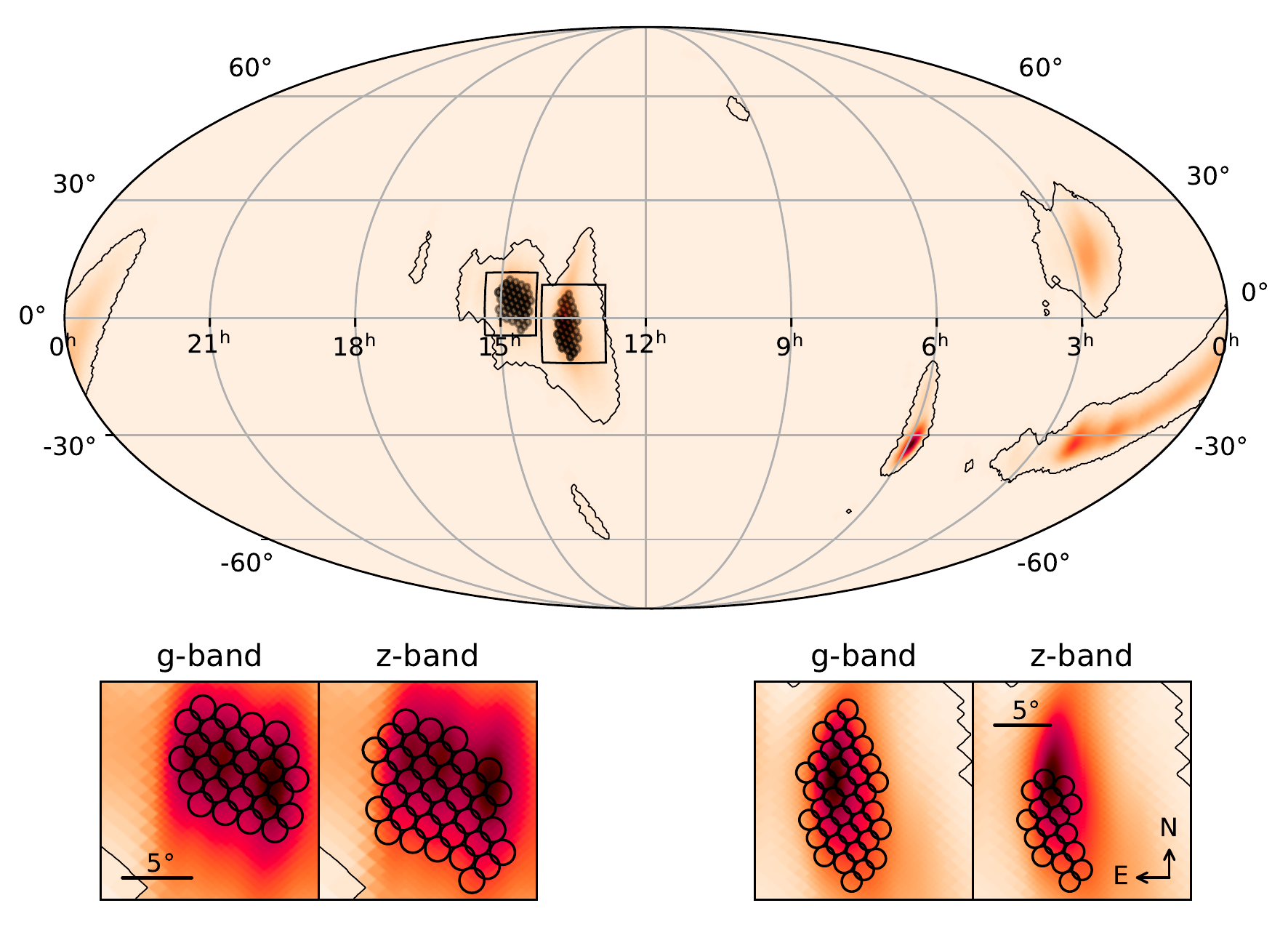}
        \includegraphics[width=0.82\textwidth]{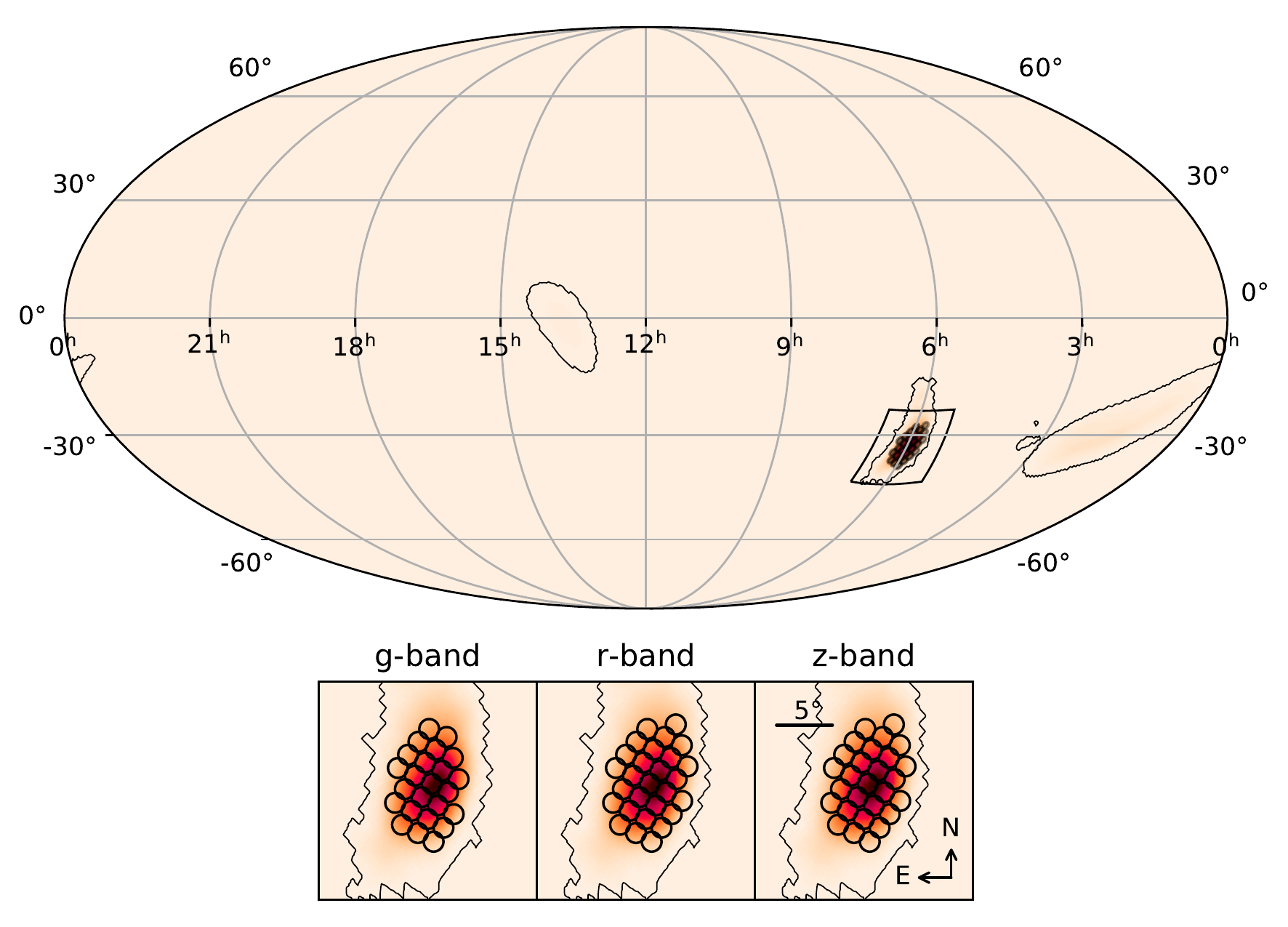}
    \caption{Preliminary BAYESTAR \citep[{\it top},][]{gcnS190510gdiscovery} and refined LALInference \citep[{\it bottom},][]{gcnS190510grefined} localization probability skymaps of the GW event S190510g. Circles show the DECam coverage, approximating the DECam FoV using a radius of 0.9\,deg$^{2}$.  
    We based our follow-up on the BAYESTAR map on the first night and the LALInference map on the second night, when it became available.  
    The high-probability patch at RA$\sim 6$\,h, where most of the probability of the LALInference skymap lies, had already set in Chile when the GW event occurred on the first  night.}
    \label{fig:location}
\end{figure*}

However, GW170817 was far better localized, and much closer to Earth, than any of the three neutron star (NS)-bearing compact binary mergers that have been detected in GWs since \citep{gcn24168,gcn24237,gcn24442}. 
The newest events have had typical distances of a few hundred Mpc, and typical localizations of about $10^3$ deg$^2$, orders of magnitude larger than those of GW170817.
Detecting the ``kilonovae" (optical and infrared transients with evolution timescales of hours to days) associated with BNS mergers at relatively large distances using a galaxy-targeted approach is challenging, as the signal is expected to be dim and galaxy catalogs are incomplete \citep{Cook2017CLU}. 
For these events, telescopes with large apertures and wide fields-of-view are required to systematically search for counterparts over large areas of sky.
An instrument well suited to this task in the Southern Hemisphere is the Dark Energy Camera \citep[DECam;][]{Flaugher2015}, a $\sim$3\,deg$^2$ wide-field imager mounted at the prime focus of the 4m Blanco telescope at the Cerro Tololo Inter-American Observatory (CTIO). 

A key challenge of the wide-field approach to GW follow-up  is determining the optimal sequence of observations to maximize counterpart discovery potential over wide areas of sky. 
Several different variables, including reference coverage, filter choice, observability, Galactic extinction, event localization, and exposure times must all be taken into account.
DECam has been extensively used in the past to follow up GW events \citep{Soares-Santos2016,Cowperthwaite2016,Annis2016,Doctor2019}, with the detection of the counterpart to GW170817 being a particular success \citep{Soares-Santos2017,Cowperthwaite2017}. 
The strategy adopted during those follow-up campaigns \citep[described in][]{Soares-Santos2016} relied on an all-sky mapping of observational parameters from models of sky brightness, atmospheric transmission, interstellar dust extinction, expected seeing, and confusion-limit probability to estimate the probability that putative GW counterparts would be detected by DECam assuming theoretical predictions for the peak luminosity. This information was then combined with the GW skymap to determine the patches of sky to be observed.

In this Letter, we describe an automated approach we have developed to solving the wide-field tiling problem, presenting a real-world application to DECam follow-up observations of the GW event S190510g, carried out as part of the DECam-GROWTH component of the Global Relay of Observatories Watching Transients Happen (GROWTH) collaboration.
In Section \ref{sec: S190510g}, we describe the GW event and its properties. 
We describe our tiling algorithm, observations,  and  data analysis methods in Section\,\ref{sec: methods}. 
In Section\,\ref{sec: results}, we present 13 high-priority transients discovered during the real-time analysis that were promptly reported to the community via the Gamma-ray Coordinate Network (GCN). In addition, we present additional 10 lower-priority transient candidates for completeness. 
In Section\,\ref{sec: discussion} we discuss both the performance of our scheduling methods and the outcome of our searches.

\section{The Gravitational Wave Event S190510g}
\label{sec: S190510g}

On 2019-05-10 02:59:39.292 UT, the compact binary merger  candidate S190510g was discovered with the Advanced LIGO and Virgo detectors in triple coincidence \citep{gcnS190510gdiscovery} using the GstLAL analysis pipeline \citep{Messick2017}. 
The event was reported to have a false alarm rate of about one in 37 years, and was initially classified as a BNS merger with 98\% probability. 
The first localization sky map obtained with the BAYESTAR software \citep{2016PhRvD..93b4013S}, released on 2019-05-10 04:03:43 UT, constrained the 90\% localization probability to a  sky area of 3462 deg$^2$. 
The luminosity distance was reported to be $269 \pm 108$\,Mpc. 
Based on S190510g's initially high probability of being a BNS merger, we triggered DECam follow-up to search for an optical counterpart. 
The observations were taken over two consecutive nights, and are described in detail in Section \ref{sec: observations}.
The initial and refined skymaps are shown in Figure\,\ref{fig:location}, along with our coverage of the event with DECam (Section \ref{sec: observations}).

A refined skymap from the LALInference localization pipeline \citep{2015PhRvD..91d2003V} was made available on 2019-05-10 10:06:59 UT, between our first and second nights of observations, reducing the 90\% probability region to a sky area of 1166 deg$^2$, with 50\% of the integrated probability enclosed in a region only 31\,deg$^2$ wide and observable with DECam \citep{gcnS190510grefined}. The modified luminosity distance was $227 \pm 92$\,Mpc. 

On 2019-05-11 20:19:22 UT, after our second and final night of observations had concluded, the LIGO Scientific Collaboration and the Virgo Collaboration (LVC) announced an update on the significance of the S190510g event \citep{gcnS190510gUpdatedClass}. The resulting re-estimate of the background model yielded a false alarm rate  of 1 in 3.6 years, significantly higher than the rate  of 1 in 37 years initially reported in \citep{gcnS190510gdiscovery}. 
The LVC reclassified S190510g to have a 58\% probability of being terrestrial and a  42\% of being a BNS, increasing the odds that this GW event was not astrophysical.
\begin{table*}
    \centering
    \begin{tabular}{cccccc| ccc | cc}
    \hline \hline
     & time start & time end & filter & $n_{exp}$ & $t_{exp}$ & prob & area & $t_{tot}$ & $\epsilon_{20}$ & $\epsilon_{30}$\\
    & UT & UT & & & [s] & [\%] & [deg$^2$] & [min] &&\\
    \hline
    night 1 & 2019-05-10 06:00:25 & 2019-05-10 07:10:02& $g$ & 65 & 30 & 15.18 & 174.20 & 70.0 & 0.78 & 0.93 \\ 
    night 1 & 2019-05-10 07:12:12 & 2019-05-10 08:22:40 &  $z$ & 54 & 30 & 10.23 & 144.72 & 70.5 & 0.64 & 0.77 \\
    night 2 & 2019-05-10 22:51:57 & 2019-05-10 23:25:50 &  $z$ & 56 & 40 & 65.02 & 75.04 & 33.9 & 0.83 & 0.96 \\
    night 2 & 2019-05-10 23:27:09& 2019-05-11 00:02:10 &  $r$ & 56 & 40 & 65.02 & 75.04 & 35.0 & 0.80 & 0.93 \\
    night 2 & 2019-05-11 00:03:27 & 2019-05-11 00:31:40 &  $g$ & 24 & 40 & 62.63 & 64.32 & 28.2 & 0.85 & 0.99 \\
    \hline
    \end{tabular}
    \caption{Summary of Observations.  We calculate the efficiency for each band using Equation\,\ref{eq: efficiency}. The overall efficiency of each observing night, calculated with $t_{tot}$ being the time difference between the last and the first exposure of the ToO observations, amounts to $\epsilon= 0.70$ on night~1 and $\epsilon = 0.80$ on night~2.}
    \label{table:summary_stats}
\end{table*}

\section{Tiling Strategy and Observations}
\label{sec: methods}

Tiling, time-allocation, and telescope scheduling were handled through \textup{gwemopt}\footnote{\url{https://github.com/mcoughlin/gwemopt}} \citep{Coughlin2018opt}, a code designed for scheduling observations of  GW skymaps with wide-field imagers.  
The code divides the GW healpix skymap into ``tiles" of the size of the field-of-view of DECam, approximated to circles with 0.9\,deg radii.  The code then determines the available segments for observation of each tile throughout the night, and applies a scheduling algorithm to the tiles to generate the order for observation.  The user can select among several scheduling algorithms while planning observations.  For each designed plan, the code calculates and displays summary statistics featuring the expected probability coverage, area tiled, and total time spent observing for each event.  We use this information to evaluate the performance of different plans in covering the accessible sky-error region.

\subsection{Scheduling Algorithms}
\label{subsec: algorithm}

\begin{figure*}
    \centering
    \includegraphics[width=0.45\textwidth]{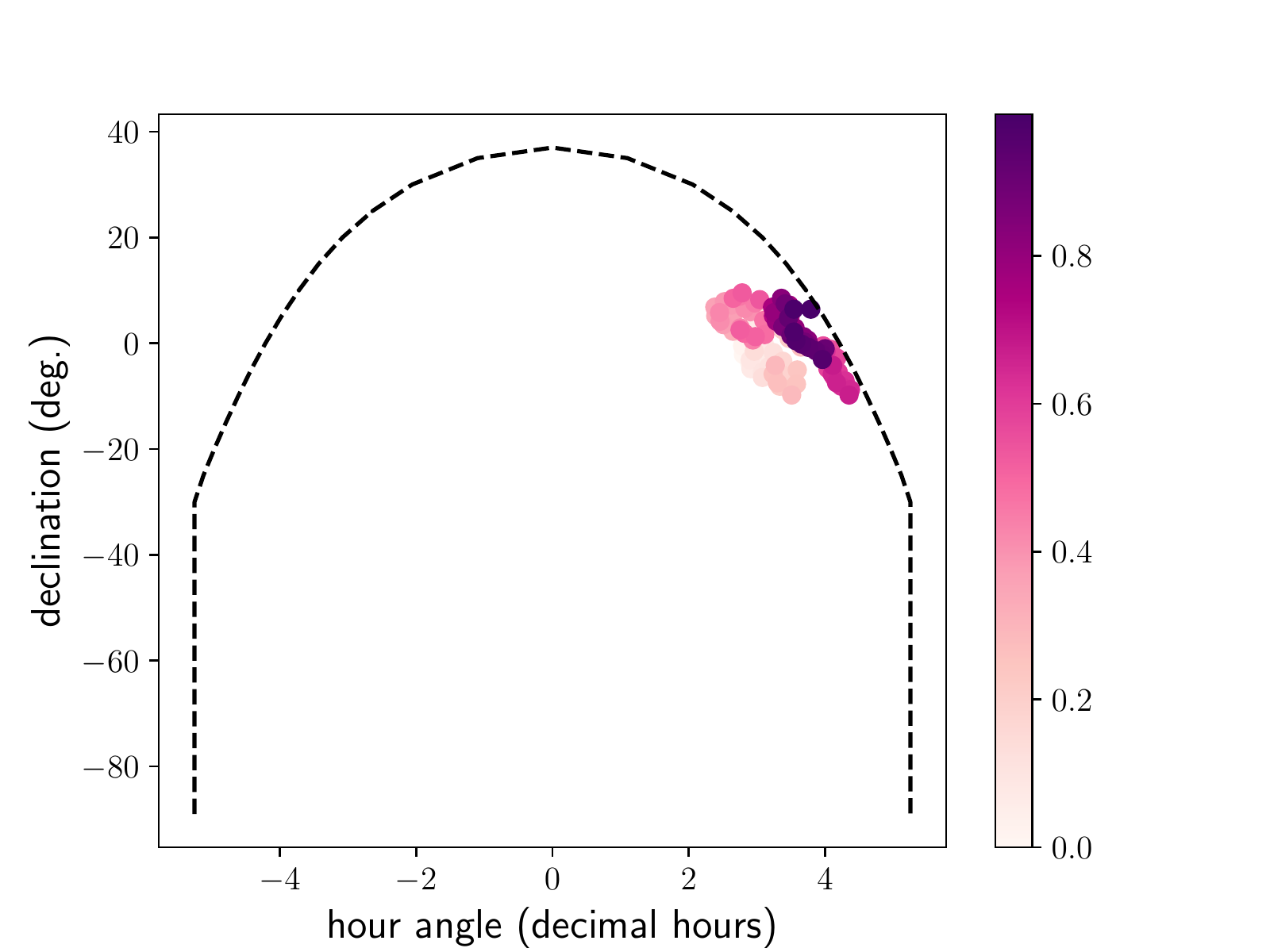}
    \includegraphics[width=0.45\textwidth]{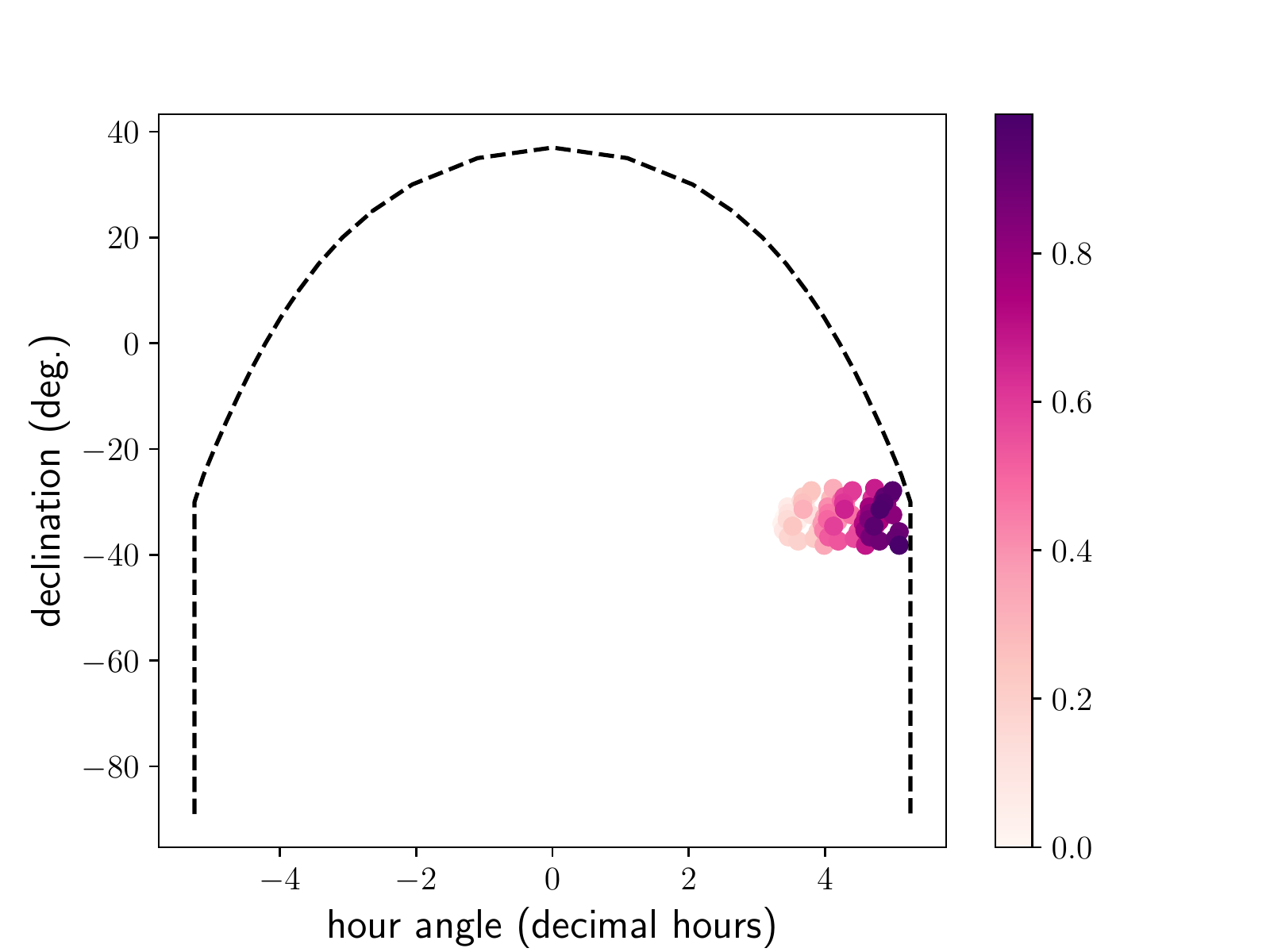}
    \caption{Hour angle as a function of declination.  These plots show the performance of the greedy-slew algorithm in scheduling observations on the first and second night within the hour-angle constraint for the Blanco Telescope (CTIO).  The colorbar, from light to dark, indicates the chronological sequence of exposures taken. Due to an unanticipated system failure and a choice of overhead times too tight to accommodate such unexpected delays, our schedule hit the hour angle limit during the first night of observations.  On the second night, with overheads optimally estimated, we could conduct 3 epochs of observations before pushing the hour angle limit.}
    \label{fig:HA_dec}
\end{figure*}

The ``greedy" algorithm is the default algorithm used for scheduling observations with gwemopt for most of the  telescopes used by GROWTH including DECam and the Zwicky Transient Facility \citep[ZTF]{2019PASP..131a8002B}.  
The greedy algorithm selects the highest probability tile that is available for observation within a given time window while taking into account setting constraints \citep{Rana2017}. While particularly effective for small-mount telescopes, the greedy algorithm can show significant limitations when long slews or slow filter changes are expected \citep{Rana2019}.
The main constraints that the Blanco telescope engineering imposes to the scheduler are a slewing rate of $\sim 1$\,deg\,s${^{-1}}$ and a firm limit on the observable hour angle as a function of the declination (see Figure \ref{fig:HA_dec}).  

As a result, we developed the ``greedy-slew" algorithm, optimized for Blanco/DECam observations. The greedy-slew algorithm is based on the greedy algorithm, but in addition it takes the Blanco hour-angle constraints into account and it penalizes large slews.  Instead of selecting a tile purely based on the probability criteria, the greedy-slew algorithm weights the probabilities based on the ratio between the readout and slew time, so that higher probability tiles with smaller slews are assigned a higher priority. 
Further optimization of the greedy-slew algorithm is in progress.

\subsection{Baseline Observing Strategy}
\label{subsec: baseline strategy}

The baseline strategy for BNS merger follow-up for our program was designed for events placed at $120 \pm 30$\,Mpc distance, the nominal angle-averaged horizon, or BNS range, of the GW detectors network during O3 \citep{Abbott2018prospects}. 
The baseline strategy consists of $g$-$z$-$g$ blocks of observations on the first night after the GW trigger, followed by $g$-$z$ blocks on the second night after the merger. 
Planned exposure times are 15\,s in $g$ and 25\,s in $z$ and sky regions where template images are available are preferred to those without pre-imaging. The exposure times were chosen based on GW170817-like kilonova models \citep{Barnes2016}. This baseline strategy was designed to rapidly identify optical counterparts via the measurement of the intra-night and inter-night color evolution of kilonovae that are expected to evolve at much faster rate than supernovae \citep[see for example][]{Kilpatrick2017Sci, Shappee2017}.

\subsection{Observations}
\label{sec: observations}

Our follow-up observations of S190510g were performed under the National Optical Astronomy Observatory (NOAO) proposal ID 2019A-0205 (PIs Goldstein and Andreoni). 
We announced the start of the observations and the availability of public DECam data via GCN \citep{gcn24443}.

We planned DECam observations using the GROWTH target-of-opportunity (ToO) Marshal\footnote{\url{https://github.com/growth-astro/growth-too-marshal}} \citep{Coughlin2019sGRB, Kasliwal2019marshal}, an open source web platform developed by the GROWTH team for the follow-up of multi-messenger transients including GW events, neutrino events, and short gamma-ray bursts. The GROWTH ToO Marshal ingests multi-messenger triggers and allows the user to plan observations for 5 facilities (DECam, ZTF, Palomar Gattini-IR, the Kitt Peak Electron Multiplying CCD demonstrator, and the GROWTH India Telescope) using programmatic follow-up algorithms.
We planned the observations described in this Letter using the greedy-slew algorithm described in Section\,\ref{subsec: algorithm}.

On the first night (hereafter ``night~1") the BAYESTAR skymap indicated that the 90\% and 50\% probability regions were 3462\,\degsq and 575\,\degsq respectively.  We started acquiring photons on 2019-10 06:00:25, or 3.01\,h after the merger, performing a block of $g$-band followed by a block of $z$-band observations, using 30\,s exposures in each band. 
Assuming an effective field of view of 2.68\,\degsq that accounts for the chip gaps, we covered 174.20\,\degsq in $g$ and 144.72\,\degsq in $z$ bands, covering 15.18\% and 10.23\% integrated probability respectively in each band. We chose against adopting a dithering pattern to cover the chip gaps in favor of a larger sky coverage.

After our night~1 observations were completed, a new LALInference skymap moved the highest probability region away from the part of the sky observed on night~1, reducing the covered integrated probability to only $\sim$2\%. As Figure \ref{fig:location} shows, the new skymap instead favored a bulge located at RA$\sim 6$\,h, Dec$\sim -35$\,deg \citep{gcnS190510grefined}. This high probability region of sky had already set at CTIO when the merger occurred on night~1, so we could not use DECam to acquire early follow-up data.

The refined skymap constrained the highest probability region (90\% and 50\% integrated probability being included in 1166\,\degsq and 31\,\degsq respectively) to be visible for $\sim 1.6$\,h at the beginning on the Chilean night on 2019-05-10 UT. 
The template coverage in the sky area with top 50\% priority was $\sim 90\%$ complete in all filters from Dark Energy Survey (DES) Data Release 1 (DR1) pre-imaging.
On the second night (hereafter ``night~2") we commenced observations on 2019-05-10 22:51:57 and finished on 2019-05-11 00:31:40, when the field set. 
We performed blocks of observations in $z$-$r$-$g$ bands using 40\,s exposures in each band.
We observed 75.04 \degsq of effective sky area in $z$ and $r$, covering 65\% of the skymap integrated probability in each band.  We then observed 64.32 \degsq in $g$ band, covering 62.3\% of the skymap integrated probability.

Other viable observing options for night~2 (that we did not select) included: i) observing a larger sky area in 2 filters with the same exposure time; ii) increasing the exposure time for at least 1 filter; iii) performing a second pass in a certain band of the observed tiles; iv) using $i$-band instead of $g$-band for the third block of observations. We excluded i) because spending the last 28\,min of observations imaging lower-probability region would have yielded only $\sim 2\%$ additional integrated probability.  In order to increase depth in $z$-band, the less sensitive among the preferred filters, we considered options ii) and iii), either with longer exposures or via image stacking of multiple exposures.  We note that 2 $z$-band epochs acquired on night~2 less than 2\,h apart were not expected to return useful information about the transient evolution, unlike $g$-band epochs pairs a few hours after the merger. The last option iv) involving $i$-band exposures was disregarded in favor of $g$-$r$-$z$ observations in order to get a more solid handle on the color of the discovered sources by covering a broader range of the optical spectrum. Assuming that a transient was detected in 2 of the chosen filters, even a low-significance detection or a non-detection in the third filter could be used as a metric to flag the source as a kilonova candidate based on existing models.  Such information is important to prioritize spectroscopic follow-up with large telescopes. 

No more DECam observations were planned on the following night(s). The total integrated probability of the LALInference skymap that we covered on the 2 observing nights with DECam is 67\%.  The observations we report here are also summarized in Table\,\ref{table:summary_stats}.

\begin{figure}[h]
    \centering
        \includegraphics[width=0.45\textwidth]{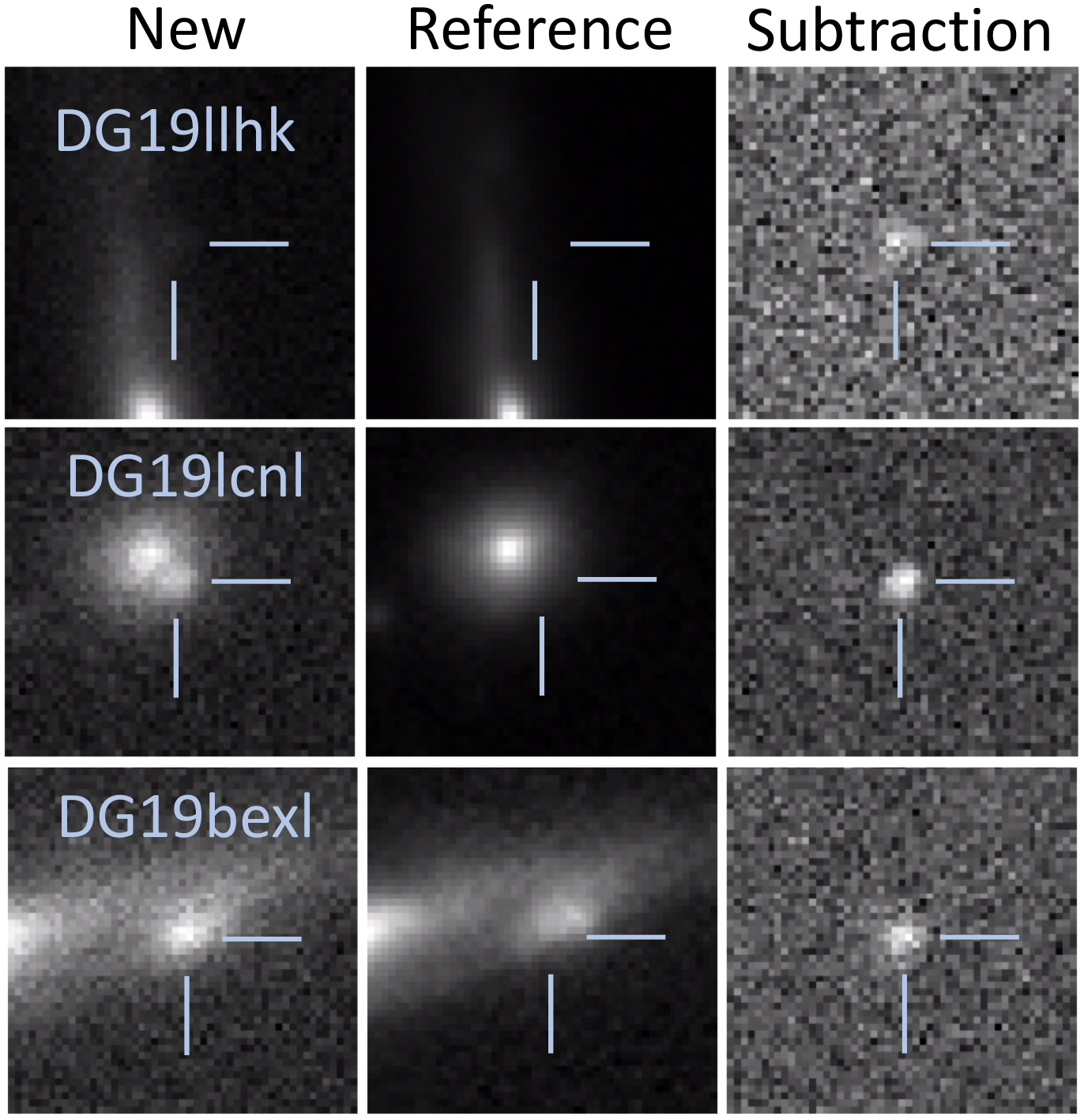}
    \caption{New image, reference image, and image subtraction of some transient candidates discovered with our image-subtraction pipeline \citep{Goldstein2019S190426cArXiv}. The side of each squared ``postage stamp" measures 13.2\,arcsec. The complete poll of candidates selected by our program during the follow-up of S190510g is presented in Section\,\ref{subsec: candidates} and in Table\,\ref{tab: candidates}--\ref{tab: candidates low-priority}.  }
    \label{fig:stamps}
\end{figure}

\begin{table*}[ht]
    \centering
    \begin{tabular}{l l c c c c c }
    \hline \hline
Name & IAU Name & RA & Dec & Date (UT) & Filter & Magnitude (AB) \\
    \hline 
DG19lcnl & AT2019fln  &$ 87.146903 $&$ -35.994405 $& 2019-05-10 23:11:24 &$ z $&$ 19.862 \pm 0.039$ \\ 
 &   & & & 2019-05-10 23:46:41&$ r $&$ 19.511 \pm 0.057$ \\ 
 &   & & & 2019-05-11 00:22:47 &$ g $&$ 20.218 \pm 0.028$ \\ 
\hline
DG19ukvo & AT2019flo  &$ 89.211464 $&$ -33.442484 $& 2019-05-10 22:51:56 &$ z $&$ > 20.51 $ \\ 
 &   & & & 2019-05-10 23:27:08 &$ r $&$ 21.338 \pm 0.049$ \\ 
 &   & & & 2019-05-11 00:03:27 &$ g $&$ 20.354 \pm 0.121$ \\ 
\hline
DG19nanl & AT2019flp  &$ 87.311394 $&$ -35.955868 $& 2019-05-10 23:11:24 &$ z $&$ 20.872 \pm 0.113$ \\ 
 &   & & & 2019-05-10 23:46:41 &$ r $&$ 19.987 \pm 0.018$ \\ 
 &   & & & 2019-05-11 00:22:47 &$ g $&$ 20.400 \pm 0.031$ \\ 
\hline
DG19qcso & AT2019flq  &$ 88.208667 $&$ -30.381390 $& 2019-05-10 23:01:23 &$ z $&$ >20.92 $ \\ 
 &   & & & 2019-05-10 23:36:39 &$ r $&$ 22.284 \pm 0.135$ \\ 
 &   & & & 2019-05-11 00:12:56 &$ g $&$ 21.545 \pm 0.087$ \\ 
\hline
DG19zaxn & AT2019flr  &$ 92.307956 $&$ -35.149825 $& 2019-05-10 23:00:07 &$ z $&$ > 20.71 $ \\ 
 &   & & & 2019-05-10 23:35:20 &$ r $&$ 20.835 \pm 0.034$ \\ 
 &   & & & 2019-05-11 00:11:37 &$ g $&$ 20.791 \pm 0.039$ \\ 
\hline
DG19etsk &  AT2019fls &$ 89.100929 $&$ -30.473990 $& 2019-05-10 23:04:59 &$ z $&$ 20.900 \pm 0.126$ \\ 
 &   & & & 2019-05-10 23:40:18 &$ r $&$ 20.712 \pm 0.036$ \\ 
 &   & & & 2019-05-11 00:16:32 &$ g $&$ 20.581 \pm 0.037$ \\ 
\hline
DG19yhhm & AT2019flt &$ 91.937008 $&$ -30.824789 $& 2019-05-10 23:16:54 &$ z $&$ > 21.64 $ \\ 
 &   & & & 2019-05-10 23:53:56 &$ r $&$ 20.080 \pm 0.019$ \\ 
 &   & & & 2019-05-11 00:29:10 &$ g $&$ 20.117 \pm 0.023$ \\ 
\hline
DG19llhk & AT2019flu &$ 90.863155 $&$ -32.385517 $& 2019-05-10 23:02:36 &$ z $&$ 20.826 \pm 0.097$ \\ 
 &   & & & 2019-05-10 23:37:52 &$ r $&$ 21.019 \pm 0.041$ \\ 
 &   & & & 2019-05-11 00:14:08 &$ g $&$ > 21.88 $ \\ 
\hline
DG19fqqk & AT2019flv  &$ 92.851450 $&$ -36.517324 $& 2019-05-10 23:08:56 &$ z $&$ 20.425 \pm 0.054$ \\ 
 &   & & & 2019-05-10 23:44:17 &$ r $&$ 20.413 \pm 0.024$ \\ 
 &   & & & 2019-05-11 00:20:26 &$ g $&$ > 22.12 $ \\ 
\hline
DG19bexl & AT2019flw  &$ 90.453717 $&$ -28.660375 $& 2019-05-10 23:14:06 &$ z $&$ 20.975 \pm 0.096$ \\ 
 &   & & & 2019-05-10 23:49:58 &$ r $&$ 21.230 \pm 0.055$ \\ 
 &   & & & 2019-05-11 00:28:02 &$ g $&$ > 21.51 $ \\ 
\hline
DG19ootl & AT2019flx &$ 87.035642 $&$ -36.076072 $& 2019-05-10 23:11:25 &$ z $&$ 21.220 \pm 0.132$ \\ 
 &   & & & 2019-05-10 23:46:41 &$ r $&$ 21.460 \pm 0.066$ \\ 
 &   & & & 2019-05-11 00:22:47 &$ g $&$ > 21.56$ \\ 
\hline
DG19nouo & AT2019fly  &$ 92.001299 $&$ -31.669159 $& 2019-05-10 23:02:36 &$ z $&$ >21.10 $ \\ 
 &   & & & 2019-05-10 23:37:52 &$ r $&$ 21.305 \pm 0.050$ \\ 
 &   & & & 2019-05-11 00:14:09 &$ g $&$ 21.436 \pm 0.067$ \\ 
\hline
DG19oahn & AT2019flz &$ 86.335286 $&$ -26.847664 $& 2019-05-10 23:24:16 &$ z $&$ 18.956 \pm 0.017$ \\ 
 &   & & & 2019-05-11 00:00:36 &$ r $&$ 19.265 \pm 0.012$ \\ 
\hline
    \end{tabular}
    \caption{Transient candidates discovered with our automated pipeline \citep{Goldstein2019S190426cArXiv} in the public DECam data that we acquired. The prefix ``DG" of the candidate names indicates their detection within the ``DECam-GROWTH" project. The observing date corresponds to the start of the exposures. The photometric measurements are not corrected for Galactic extinction. These candidates \citep[other than DG19qcso,][]{gcn24540} were reported $\sim 3$\,h after the beginning of the observations \cite{gcn24467} and have higher priority.  All the candidates (including DG19qcso) are likely associated with prominent host galaxies.}
        \label{tab: candidates}
\end{table*}

\begin{table*}[ht]
    \centering
    \begin{tabular}{l l c c c c c }
    \hline \hline
Name & IAU Name & RA & Dec & Date (UT) & Filter & Magnitude (AB)  \\
    \hline 
DG19bann & AT2019fne  &$ 87.388444 $&$ -33.403834 $& 2019-05-10 22:56:31 &$ z $&$ > 20.19 $ \\ 
 &   & & & 2019-05-10 23:31:44 &$ r $&$ 21.397 \pm 0.060$ \\ 
 &   & & & 2019-05-11 00:08:03 &$ g $&$ 21.560 \pm 0.086$ \\ 
\hline
DG19zzwl & AT2019fnf  &$ 91.127081 $&$ -29.319263 $& 2019-05-10 23:14:06 &$ z $&$ 20.895 \pm 0.089$ \\ 
 &   & & & 2019-05-10 23:49:58 &$ r $&$ 20.680 \pm 0.031$ \\ 
 &   & & & 2019-05-11 00:25:23 &$ g $&$ 20.740 \pm 0.042$ \\ 
\hline
DG19pybq & AT2019fnh &$ 87.047680 $&$ -35.372392 $& 2019-05-10 23:11:25 &$ z $&$ > 21.32 $ \\ 
 &   & & & 2019-05-10 23:46:41 &$ r $&$ 21.750 \pm 0.084$ \\ 
 &   & & & 2019-05-11 00:22:47 &$ g $&$ 21.510 \pm 0.082$ \\ 
\hline
DG19qpqp & AT2019fni  &$ 91.092625 $&$ -29.766986 $& 2019-05-10 23:14:06 &$ z $&$ > 21.58 $ \\ 
 &   & & & 2019-05-10 23:49:58 &$ r $&$ 21.874 \pm 0.096$ \\ 
 &   & & & 2019-05-11 00:25:22 &$ g $&$ 21.795 \pm 0.116$ \\ 
\hline
DG19fbio & AT2019fnj &$ 91.565165 $&$ -31.464310 $& 2019-05-10 23:02:36 &$ z $&$ > 21.29 $ \\ 
 &   & & & 2019-05-10 23:37:52 &$ r $&$ 22.123 \pm 0.145$ \\ 
 &   & & & 2019-05-11 00:14:09 &$ g $&$ 22.086 \pm 0.120$ \\ 
\hline
DG19cgep & AT2019fnk &$ 91.832510 $&$ -33.142566 $& 2019-05-10 23:06:10 &$ z $&$ > 21.21 $ \\ 
 &   & & & 2019-05-10 23:41:28 &$ r $&$ 20.749 \pm 0.136 $ \\ 
 &   & & & 2019-05-11 00:17:43 &$ g $&$ 21.985 \pm 0.109$ \\ 
\hline
DG19dbln & AT2019fnl &$ 90.077019 $&$ -34.912666 $& 2019-05-10 22:55:22 &$ z $&$ >20.59 $ \\ 
 &   & & & 2019-05-10 23:30:35 &$ r $&$ 21.773 \pm 0.113$ \\ 
 &   & & & 2019-05-11 00:06:53 &$ g $&$ 21.647 \pm 0.093$ \\ 
\hline
DG19soko & AT2019fnm &$ 91.740720 $&$ -32.094698 $& 2019-05-10 23:02:35 &$ z $&$ >21.13 $ \\ 
 &   & & & 2019-05-10 23:37:52 &$ r $&$ 21.781 \pm 0.075$ \\ 
 &   & & & 2019-05-11 00:14:09 &$ g $&$ 21.864 \pm 0.100$ \\ 
\hline
DG19ujcn & AT2019fnw &$ 91.048561 $&$ -34.805273 $& 2019-05-10 22:54:13 &$ z $&$ >20.69 $ \\ 
 &   & & & 2019-05-10 23:29:27 &$ r $&$ 21.791 \pm 0.106$ \\ 
 &   & & & 2019-05-11 00:05:44 &$ g $&$ 21.512 \pm 0.076$ \\ 
\hline
DG19qoln & AT2019fnx &$ 89.570078 $&$ -35.443861 $& 2019-05-10 22:55:22 &$ z $&$ > 20.74 $ \\ 
 &   & & & 2019-05-10 23:30:35 &$ r $&$ 21.863 \pm 0.090$ \\ 
 &   & & & 2019-05-11 00:06:53 &$ g $&$ 21.921 \pm 0.113$ \\ 
\hline
    \end{tabular}
    \caption{Same as Table\,\ref{tab: candidates}, but presenting lower-priority candidates that did not meet the criteria that we described in Section\,\ref{subsec: candidates}.  The low-priority candidates are reported here for completeness.  }
        \label{tab: candidates low-priority}
\end{table*}

\subsection{Data analysis}
\label{subsec: pipeline}

We processed and analyzed images using an image subtraction pipeline we developed for the discovery of GW counterparts using DECam. The pipeline, more extensively described in \cite{Goldstein2019S190426cArXiv}, automatically transfers data from NOAO servers and processes them in parallel at the National Energy Research Scientific Computing Center (NERSC) at Lawrence Berkeley National Laboratory. Image calibration includes flat-fielding, overscan correction, and bad pixel/column masking.  The pipeline computes an astrometric solution against {\it Gaia} DR1 \citep{Gaia2016} sources using the \textup{Scamp} package.  New DECam images are aligned with reference images (or ``templates") using \textup{SWarp}.  We use the HOTPANTS \citep{Becker2015} implementation of the \cite{Alard2000} algorithm to perform image subtraction, the autoScan package to perform artifact rejection \citep{2015AJ....150...82G}, and  the \textup{PSFEx} and \textup{SExtractor} packages to perform PSF photometry.

The pipeline obtains reference images from the DECam Legacy Survey (DECaLS) Data Release 7 and the DES Data Release 1. It calibrates photometric zero points against the DECaLS and DES catalogs. 
We used the GROWTH marshal online platform \citep{Kasliwal2019marshal} to collect and vet transient candidates before announcing them via GCN circulars.

\section{Results}
\label{sec: results}

\subsection{Transient Candidates Discovered}
\label{subsec: candidates}

The automated pipeline described in Section\,\ref{subsec: pipeline} and in \cite{Goldstein2019S190426cArXiv} started yielding transient candidates $\sim 15$ minutes after the observations started at CTIO on night~1. 
The LALInference skymap \citep{gcnS190510grefined}, made available shortly after we finished the DECam observations on night~1, largely ruled out those transients as  potential counterparts to S190510g. Therefore we did not issue a GCN reporting transient discoveries on night~1 and here we present only those candidates discovered on night~2. On that night our pipeline successfully analyzed 89\% of the CCDs, with the remaining $11\%$ being not processed mainly due to the lack of reference images.  

We vetted and prioritized candidates based on their likelihood to be extragalactic sources with host galaxies within the distance range expected for S190510g. In order to reject asteroids, our criteria for selecting candidates required two detections separated in time by $>30$\,min.  We did not include nuclear sources among the highest priority candidates. We also excluded known transients already reported on the Transient Name Server, including SN2019bso, from the candidate list. 

Using the above criteria, we narrowed down the list from 176802 to 12 candidates and we reported them via GCN \citep{gcn24467} on 2019-05-11 02:18:53 UT, $\sim$3.5\,h after the start of the observations. Figure \ref{fig:stamps} shows example ``postage stamp" images of three of the candidates from the GCN.
Table\,\ref{tab: candidates} summarizes relevant information and photometry of those candidates. The table includes DG19qcso (where the prefix ``DG" indicates the detection of the candidate by the ``DECam-GROWTH" project), a transient candidate that met the above criteria, but that was reported in a separate GCN circular \citep{gcn24540}.  In Table\,\ref{tab: candidates low-priority} we report other candidates that appear to be nuclear, hostless, or with faint host galaxies likely placed at large distances for completeness. 

Two of the reported transients seemed to be of particular interest. The first candidate, DG19llhk,  appeared to be associated with a host galaxy located at redshift $z = 0.07158$ according to the 6dF galaxy survey  \citep{jones2009}. The redshift was consistent with the LVC distance estimate. At that distance, the absolute magnitude of DG19llhk was $-16.4$, consistent with GW170817 at +1 day. The second candidate of interest was DG19lcnl, whose color from preliminary photometry ($m_g =  20.27 \pm 0.04$, $m_r = 19.45 \pm 0.02$, $m_z = 20.18 \pm 0.06$) appeared consistent with GW170817 at the same phase. The lack of a coincident source in the reference image and the proximity of the transient to a galaxy (Figure~\,\ref{fig:stamps}) suggested the nature of DG19lcnl to be extragalactic, although a redshift of the putative host was not present in survey catalogs.

Photometric follow-up with the Korea Microlensing Telescope Network \citep[KMTNet,][]{Kim2016KMTNet} between 2019-05-10 16:48:24 UT and before 2019-05-11 23:42:11 UT indicated no significant fading ($\Delta R > 0.5$\,mag\,day$^{-1}$) for any of the 12 transients reported by \cite{gcn24467} (including DG19llhk and DG19lcnl) or with respect to our preliminary $r$-band DECam photometry \citep{gcn24466,gcn24493,gcn24529}. A decay of $\Delta r \gtrsim 0.5$\,mag\,day$^{-1}$ is expected from kilonovae at this phase, thus the KMTNet photometry provided evidence against any of the 12 transients in the GCN being the counterpart to S190510g. 

\cite{gcn24511} used the IMACS Spectrograph on the 6.5m Magellan-Baade telescope to spectroscopically classify the candidate DG19fqqk as a Type\,II supernova at redshift $z = 0.06$, confirming that the source was unrelated to the GW event. Observations performed after the GW trigger time, with the X--ray Telescope (XRT) onboard the {\it Neil Gehrels Swift Observatory}, revealed no X--ray source at the location of DG19fqqk with a 3$\sigma$ upper limit of $6.1 \times 10^{-2}$\,ct\,s$^{-1}$, corresponding to a 0.3--10\,keV flux of $2.6 \times 10^{-12}$\,erg\,cm$^{-2}$\,s$^{-1}$ \citep{gcn24541}. 
XRT upper limits are also available for the candidate DG19llhk, for which no X--ray flux was detected down to $6.4 \times 10^{-2}$\,ct\,s$^{-1}$ (3$\sigma$),  corresponding to $2.8 \times 10^{-12}$\,erg\,cm$^{-2}$\,s$^{-1}$ \citep{gcn24541}. 

The DES Collaboration reported 11 transients from an independent analysis the public DECam data \citep{gcn24474, gcn24480}. Six of those candidates confirmed detections already reported in \cite{gcn24467} and one corresponded to the known supernova SN2019bso.
The four remaining candidates were not reported by our team. The desgw-190510a and desgw-190510f candidates were detected only once by our pipeline (thus not passing our selection criteria), while desgw-190510b and desgw-190510g were not detected in any band. In particular, desgw-190510a was detected by our pipeline in $r$-band ($r = 20.980 \pm 0.065$), but it was not detected in $g$-band (faint, $g > 21.8$) and $z$-band (rejected by our automatic cuts on the candidate shape).
The candidate desgw-190510b had no template coverage in DES DR1.
The candidate desgw-190510g was not detected in $g$-band (rejected) and $z$-band ($z > 21.4$), with no DES DR1 coverage available in $r$-band.
Finally, desgw-190510f was detected by our pipeline in $r$-band ($r = 21.296 \pm 0.051$), but it was not detected in $g$-band (rejected) and $z$-band (faint, $z > 20.8$).

\subsection{Observing efficiency}

We define the  efficiency $\epsilon$ of our  observations as the ratio of the time spent on sky (accounting for the minimum overhead per exposure, mostly dictated by the CCD readout) over the total duration of the observations

\begin{equation}
\label{eq: efficiency}
\epsilon = \frac{ n_\mathrm{exp}  \times (\text{exptime} + \text{overhead})}{t_\mathrm{tot}}
\end{equation} 

where $n_\mathrm{exp}$ represents the number of exposures, ``overhead'' is the minimum overhead possible per exposure, and $t_\mathrm{tot}$ is the duration of the observations.  Such a definition of efficiency is also valid outside the GW follow-up context.

Assuming 20\,s of CCD readout to be the dominant overhead, we obtained an efficiency $\epsilon= 0.70$ on night~1 and $\epsilon = 0.80$ on night~2. More realistically, the overhead time for each exposure amounts to $\sim 30$\,s even without any slewing \citep[see for example][]{Andreoni2019fast}, in which case the efficiency increases to $\epsilon = 0.84$ on night~1 and $\epsilon = 0.94$ on night~2. The lower efficiency of night~1 is almost entirely due to 2 outliers in the distribution of slewing times. A CTIO computer failure caused a loss of 13.8\,min starting at 2019-05-10 08:08 UT. A 100.2\,s slew between the end of the $g$-band observing block and the beginning of the $z$-band observing block constitutes the second outlier.  
Removing those outliers, the efficiency between night~1 and night~2 is comparable within $< 2\%$, with the difference due to larger slews dictated by the larger area to cover (in other words, by the less precise localization skymap). 

The observing efficiency using the greedy-slew algorithm was superior to the greedy algorithm for DECam.  During the DECam follow-up of the GW event S190426c \citep{Goldstein2019S190426cArXiv} the overall efficiency of our observations was $\epsilon = 0.52$ and $\epsilon = 0.60$ assuming $ 20$\,s and $\sim 30$\,s overhead per exposure respectively.  The greedy-slew algorithm optimized for DECam, along with a better (more conservative) overhead time estimation, led to a significant improvement during the follow-up of S190510g.  Those new techniques were developed in $< 2$ weeks by the GROWTH ToO marshal team, after S190426c and before S190510g happened.

\begin{figure*}[t]
\centering
\includegraphics[width=1.5\columnwidth]{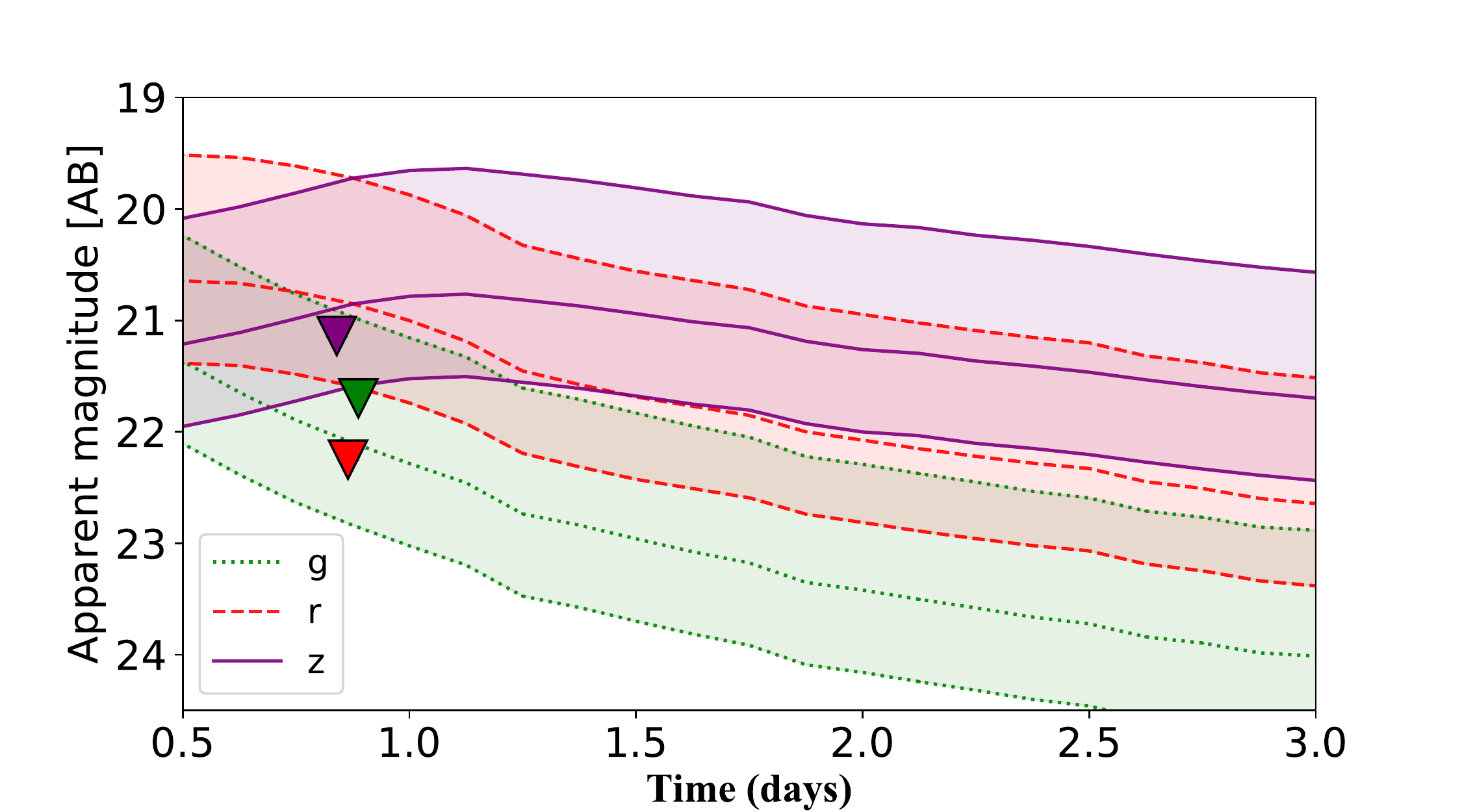}
\includegraphics[width=2.1\columnwidth]{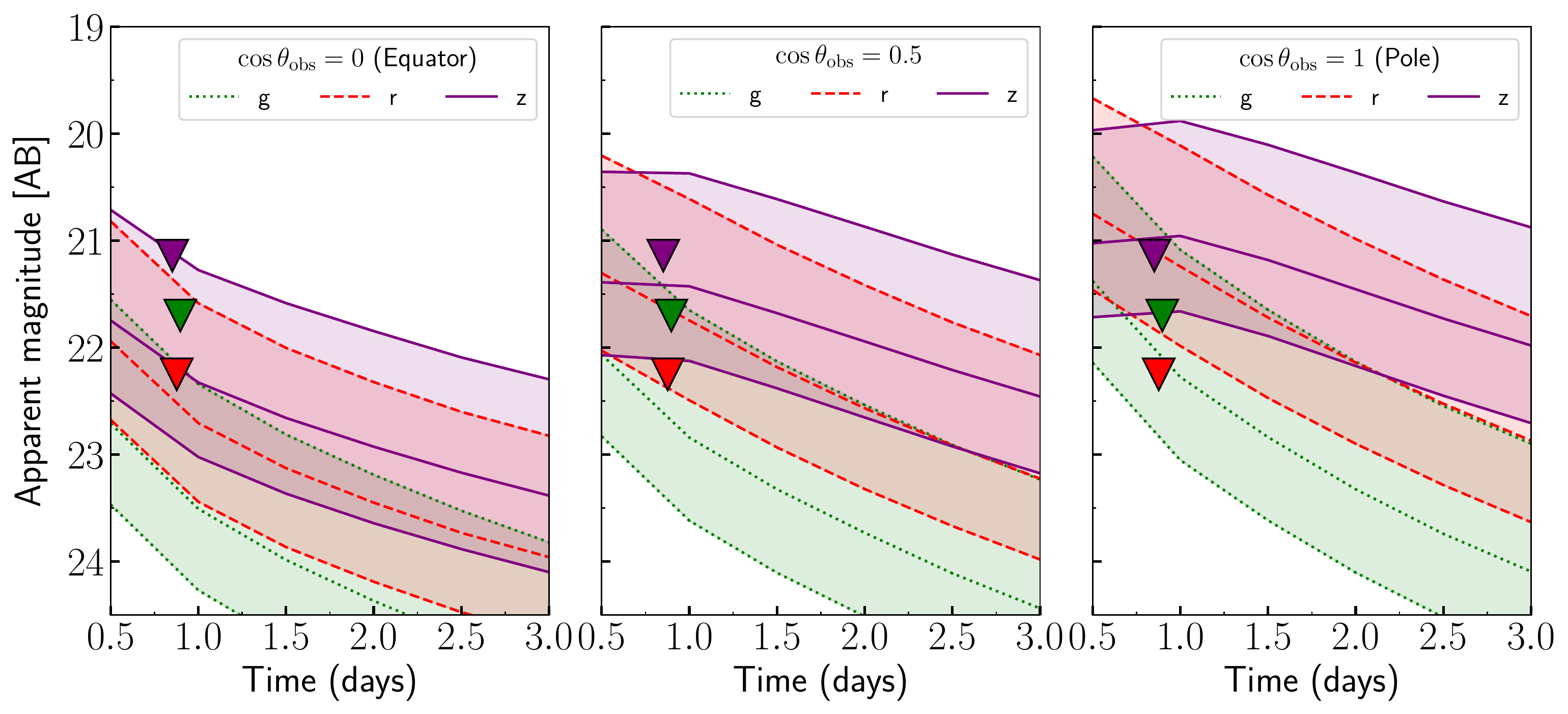}
\caption{Kilonova models developed by \cite{Kasen2017Nat} ({\it upper panel}) and \cite{Bulla2019NatAs,Bulla2019arXiv}. ({\it lower panels}).  All those models account for two kilonova components, a ``blue" lanthanide-poor and a ``red" lanthanide-rich component.  Triangles indicate median limiting magnitudes of our observations. In the {\it upper} panel, the GW170817-like kilonova is placed at the distance range expected for S190510g. In the {\it lower} panel, the models by \cite{Bulla2019NatAs, Bulla2019arXiv} show how the GW170817 kilonova emission would look like at three different viewing angles (from edge-on, $\cos\theta_\mathrm{obs}=0$, to face-on, $\cos\theta_\mathrm{obs}=1$). The extension of the lanthanide-rich component around the equatorial/merger plane is parameterized by an half-opening angle set to $\phi = 30$\,deg, which best fits the observed light curve of GW170817 \citep{Bulla2019arXiv}.    
The plots presented in this figure show that we would have likely been able to detect a kilonova with the same properties of GW170817 in at least 2 bands at the central distance of the distribution inferred from the GW data analysis under favorable viewing angles.}
\label{fig: model}
\end{figure*}

\section{Discussion}
\label{sec: discussion}

The follow-up of S190510g highlighted the trade-off between depth, area to cover, and color information that must be faced during GW follow-up of BNS mergers during O3. The event had a very large localization area on night~1 (3462 \degsq for the 90\% probability) and its initial distance was at least 3 times the distance to GW170817.  At such large distances, the expected kilonova emission may be too faint to be detected with 1m-class telescopes and galaxy-targeted follow-up becomes less efficient due to poorer galaxy catalog completeness and larger numbers of galaxies enclosed in the highest-probability volume.  

Covering large sky areas, as in the case of S190510g (and S190426c before that), is made possible by optimized algorithms (see Section\,\ref{subsec: algorithm}) that push the observations close to the physical limits of the telescope mount.  Our new greedy-slew algorithm was able to achieve a total of $\epsilon = 94\%$ efficiency and up to $\epsilon = 99\%$ excluding filter changes on the second observing night. 
If the telescope, the dome, or the observing software experiences issues that causes the observations to stop temporarily (as it happened on night~1 of the observations presented here) the schedule should be re-computed in real time.  

The follow-up of S190510g also highlighted the relevance of DECam observations when the highest probability localization region is located at Dec $< -30$\,deg, too far South for many northern hemisphere survey facilities (including, for example, ZTF, ATLAS, and Pan-STARRS) to observe. High-probability skymap regions were observable from CTIO on both night~1 and night~2, even after the skymap change.  

In Figure\,\ref{fig: model} we compare the limiting magnitudes of night~2 with a sample of kilonova models \citep{Kasen2017Nat, Bulla2019NatAs,Bulla2019arXiv}. The
3$\sigma$ limiting magnitudes of $g < 21.7$, $r <  22.3$, and $z < 21.2$ were calculated by estimating the magnitude of a source with a count rate per pixel three times higher than the sky background. DG
With 40\,s exposure time in $g$-$r$-$z$ bands, we could have detected a GW170817-like kilonova \citep[Figure\,\ref{fig: model}, upper panel, based on][]{Kasen2017Nat} in at least 2 filters out to $\sim 227$\,Mpc, the centre of the distance distribution inferred by the LALInference skymap \citep[$227 \pm 92$\,Mpc,][]{gcnS190510grefined}. 

Other models \citep[Figure\,\ref{fig: model}, lower panels, based on][]{Bulla2019NatAs, Bulla2019arXiv} allow us to explore the detectability of a GW170817-like kilonova at different viewing angles. The comparison of the models with the DECam detection limits suggest good detection chances in 2 or 3 filters out to $\sim 230$\,Mpc under polar viewing angles (i.e., when the kilonova is viewed face-on), with the identification of a counterpart becoming more difficult at equatorial viewing angles.  
We stress that the non-detection in 1 band can add important information to help reject contaminants (such as supernovae) and lead to rapid identification of the most promising kilonova candidates.  We note that the observations of GW170817 provided well-sampled, multi-wavelength light curves that we could use to better plan our follow-up campaigns and to assess the results of our kilonova searches in real time.  However, the kilonova population is likely diverse, as recent (post-GW170817) studies of short gamma-ray burst optical afterglows suggest \citep{Fong2017,Gompertz2018,Troja2018KN,Rossi2019arXiv}. The discovery of a larger population of kilonovae will allow us to get a more solid handle on the distribution of their physical and observational properties. Key to the detection of more GW optical counterparts is the combination of deep imaging and wide-area coverage, that our DECam observations aim to maximize. The upcoming Large Synoptic Survey Telescope is expected to be more sensitive than DECam and, thanks to its $\sim 10$\,deg$^2$ field of view, may allow us to probe the fainter end of the kilonova luminosity distribution even for those mergers localized over an area of hundreds of deg$^2$.

Optical counterpart candidates discovered with our automatic pipeline (Section\,\ref{subsec: pipeline}) were carefully vetted and made public $\sim 3$\,h after the beginning of the observations (Section\,\ref{sec: results}).  This enabled multi-wavelength photometric and spectroscopic follow-up to be promptly performed on candidates that we and other groups prioritized.
Candidates could be made available in batches every $\sim 2$\,h during future follow-ups.

\section{Conclusion}
\label{sec: conclusion}

In this paper we presented our DECam follow-up of S190510g, a possible binary neutron star merger discovered by LVC. The observations were planned using the GROWTH ToO marshal scheduling platform \citep{Coughlin2019sGRB, Kasliwal2019marshal} using algorithms that we have optimized specifically for DECam, but that could be applied to other facilities. 
On the second night of our DECam follow-up of S190510g we covered 65\% of the localization probability.  We reached a total observing efficiency of $\epsilon = 0.94$, which was a marked improvement over our first DECam GW follow-up of O3 \citep[$\epsilon = 0.60$,][]{Goldstein2019S190426cArXiv}.

Our observations were deeper than planned in our baseline strategy to account for the distance to the event larger than expected, possibly as large as $\sim 8$ times the distance to GW170817. We estimate that our observations can lead to the detection of GW170817-like kilonovae in at least 2 bands at $\sim 230$\,Mpc over an area of 75\,\degsq in the 1.66\,h of observing time that was available, albeit under favorable viewing angles. 

Additional information such as constraints on the viewing angle 
from early multi-messenger analysis could help in choosing the most appropriate combination of filters, cadence, and exposure times for future electromagnetic follow-ups \citep[see for example][]{Chen2018arXiv}. Coarsely localized GW events at (or beyond) the angle-averaged horizon of the Advanced LIGO and Virgo detectors may be common during O3, thus optimally scheduled observations with DECam can be key in upcoming follow-up of new GW triggers.

\acknowledgements
\section*{Acknowledgments}
On behalf of the entire GROWTH team, I.A. and D.A.G. gratefully acknowledge Steve Heathcote, Kathy Vivas, Tim Abbott, and the staff at CTIO and NOAO for facilitating these target of opportunity observations. 
I.A. and M.M.K. thank Myungshin Im and the KMTNet team for coordinating  follow-up observations of DECam-GROWTH candidates. The authors thank the anonymous referee for the useful review of the manuscript.  

This work was supported by the GROWTH (Global Relay of Observatories Watching Transients Happen) project funded by the National Science Foundation under PIRE Grant No 1545949. GROWTH is a collaborative project among California Institute of Technology (USA), University of Maryland College Park (USA), University of Wisconsin Milwaukee (USA), Texas Tech University (USA), San Diego State University (USA), University of Washington (USA), Los Alamos National Laboratory (USA), Tokyo Institute of Technology (Japan), National Central University (Taiwan), Indian Institute of Astrophysics (India), Indian Institute of Technology Bombay (India), Weizmann Institute of Science (Israel), The Oskar Klein Centre at Stockholm University (Sweden), Humboldt University (Germany), Liverpool John Moores University (UK) and University of Sydney (Australia).

\noindent D.A.G. acknowledges support from Hubble Fellowship grant HST-HF2-51408.001-A.
S.A. acknowledges support from the PMA Division Medberry Fellowship at the California Institute of Technology.
Support for Program number HST-HF2-51408.001-A is provided by NASA
through a grant from the Space Telescope Science Institute, which is operated by the Association of Universities for Research in Astronomy, Incorporated, under NASA contract NAS5-26555.
M.W.C. is supported by the David and Ellen Lee Postdoctoral Fellowship at the California Institute of Technology.
M.B. and E.K. acknowledge support from the G.R.E.A.T. research environment funded by the Swedish National Science Foundation.
P.N. acknowledges support from the DOE through DE-FOA-0001088, Analytical Modeling for Extreme-Scale Computing Environments. 
J.S.B., J.M.-P., and K.Z. are partially supported by a Gordon and Betty Moore Foundation Data-Driven Discovery grant. 
Support for J.B. was provided in part by the National Aeronautics and Space Administration(NASA)through the Einstein Fellowship Program, grant number PF7-180162.
P.G. is supported by NASA Earth and Space Science Fellowship (ASTRO18F-0085).
A.A.M. acknowledges support from the following grants: NSF AST-1749235, NSF-1640818 and NASA 16-ADAP16-0232.
V.Z.G. acknowledges support from the University of Washington College of Arts and Sciences, Department of Astronomy, and the DIRAC Institute. University of Washington's DIRAC Institute is supported through generous gifts from the Charles and Lisa Simonyi Fund for Arts and Sciences, and the Washington Research Foundation. 
A.K.H.K. acknowledges support from the Ministry of Science and Technology of the Republic of China (Taiwan) under grants 106-2628-M-007-005 and 107-2628-M-007-003.

\noindent This project used data obtained with the Dark Energy Camera (DECam), which was constructed by the Dark Energy Survey (DES) collaborating institutions: Argonne National Lab, University of California Santa Cruz, University of Cambridge, Centro de Investigaciones Energeticas, Medioambientales y Tecnologicas-Madrid, University of Chicago, University College London, DES-Brazil consortium, University of Edinburgh, ETH-Zurich, University of Illinois at Urbana-Champaign, Institut de Ciencies de l'Espai, Institut de Fisica d'Altes Energies, Lawrence Berkeley National Lab, Ludwig-Maximilians Universitat, University of Michigan, National Optical Astronomy Observatory, University of Nottingham, Ohio State University, University of Pennsylvania, University of Portsmouth, SLAC National Lab, Stanford University, University of Sussex, and Texas A$\&$M University. Funding for DES, including DECam, has been provided by the U.S. Department of Energy, National Science Foundation, Ministry of Education and Science (Spain), Science and Technology Facilities Council (UK), Higher Education Funding Council (England), National Center for Supercomputing Applications, Kavli Institute for Cosmological Physics, Financiadora de Estudos e Projetos, Funda\c{c}\~{a}o Carlos Chagas Filho de Amparo a Pesquisa, Conselho Nacional de Desenvolvimento Cient\'{i}fico e Tecnol\'{o}gico and the Minist\'{e}rio da Ci\^{e}ncia e Tecnologia (Brazil), the German Research Foundation-sponsored cluster of excellence ``Origin and Structure of the Universe" and the DES collaborating institutions.

\noindent This work has made use of data from the European Space Agency (ESA) mission {\it Gaia} \citep{Gaia2016,Gaia2018} (\url{https://www.cosmos.esa.int/gaia}), processed by the {\it Gaia} Data Processing and Analysis Consortium (DPAC,
\url{https://www.cosmos.esa.int/web/gaia/dpac/consortium}). Funding for the DPAC has been provided by national institutions, in particular the institutions participating in the {\it Gaia} Multilateral Agreement.

\bibliography{references}

\end{document}